\documentclass[preprint,pra,showpacs,nofootinbib]{revtex4}
\usepackage[mathscr]{euscript}
\usepackage{dcolumn}
\usepackage{bm}
\usepackage{graphicx}
\usepackage{dcolumn}
\usepackage{amsmath,amssymb,amsthm}
\usepackage[colorlinks=true,linkcolor=blue]{hyperref}
\newcommand{\bea}{\begin{eqnarray}}
\newcommand{\eea}{\end{eqnarray}}
\newcommand{\beq}{\begin{equation}}
\newcommand{\eeq}{\end{equation}}
\newcommand{\nn}{\nonumber}
\def\/{\over}

\begin{document}

\title{Geometric phase for an accelerated two-level atom and the Unruh effect}
\author{  Jiawei Hu$^{1}$ and Hongwei Yu$^{1,2,}$\footnote{Corresponding author} }
\affiliation{$^1$ Institute of Physics and Key Laboratory of Low
Dimensional Quantum Structures and Quantum
Control of Ministry of Education,\\
Hunan Normal University, Changsha, Hunan 410081, China \\
$^2$ Center for Nonlinear Science and Department of Physics, Ningbo
University, Ningbo, Zhejiang 315211, China}


\begin{abstract}
We study, in the framework of open quantum systems, the geometric
phase acquired by a uniformly accelerated two-level atom undergoing
nonunitary evolution due to its coupling to a bath of fluctuating
vacuum electromagnetic fields in the multipolar scheme. We find that
the phase variation due to the acceleration can be in principle
observed  via atomic interferometry between the accelerated atom and
the inertial one, thus providing an evidence of the Unruh effect.

\end{abstract}

\maketitle


When a quantum system undergoes a cyclic evolution, it may acquire a
memory of this motion in the form of a geometric phase. This phase
was first introduced by Pancharatnam while studying polarized beams
passing through crystals~\cite{Pancharatnam}. In 1984, Berry studied
the dynamics of a closed quantum system whose Hamiltonian varies
adiabatically in a cyclic way, and found, besides the familiar
dynamical phase, that there is an additional phase due to the
geometry of the path enclosed during the evolution of the system in
the parameter space~\cite{berry}. Berry's work was soon generalized
to nonadiabatic~\cite{nonadiabatic} and noncyclic
evolution~\cite{noncyclic}. The geometric phase has so far been
extensively studied, both theoretically and
experimentally~\cite{GPbook}, and it has been fruitfully applied to
many fields, such as the study of molecular
dynamics~\cite{molecular} and electronic properties~\cite{solid}.

Recently, there has been interest in using the geometric phase for
fault-tolerant quantum computation~\cite{fault-tolerant}. However,
due to the inevitable interactions between the qubits and the
environment, a pure state will be driven to a mixed state under the
environment induced decoherence and dissipation. As a result, the
geometric phase has to be generalized to general evolutions of an
open system. Uhlmann was the first to define a mixed-state geometric
phase via mathematical concept of purification~\cite{Uhlmann}.
Sj\"{o}qvist et al. put forward an alternative definition for the
unitarily evolved nondegenerate mixed-state density matrix based on
the interferometry~\cite{Sjoqvist}. This  was soon generalized to
degenerate mixed states by Singh et al.~\cite{Singh} and to the
nonunitary evolution using the kinematic approach by Tong et
al.~\cite{Tong}.
Wang et al. defined a mixed-state geometric phase via mapping the
density matrix to a nonunit vector ray in the complex projective
Hilbert space \cite{Wang}. Experiments based on NMR system~\cite{Du}
and single photon interferometry~\cite{Marie} have demonstrated the
mixed-state geometric phase.

As discussed above, the impact of environment on the geometric phase
of open systems is an important issue in any practical
implementations of quantum computing. In this regard, the effects of
different kinds of decoherence sources on the geometric phase, such
as dephasing and spontaneous decay, haven been
analyzed~\cite{Carollo}. In Ref~\cite{Rezakhani}, Rezakhani et al.
have studied geometric phase for an open system, which is a
spin-half particle in weak coupling to a thermal bath, and found
that the phase varies with the temperature of the bath. Lombardo et
al.~\cite{Lombardo} have studied not only how the geometric phase is
modified by the presence of the different types of environments, but
also estimated the corresponding times at which decoherence becomes
effective. Chen et al. \cite{chen} focused on the geometric phase of
an open two-level atom coupled to an environment with Lorentzian
spectral density and explored the non-Markovian effect on the
geometric phase.


In quantum sense,  every system, whatever it is, is an open system,
since it is at least subjected to vacuum fluctuations. However, the
geometric phase of an open system generated by the nonunitary
evolution due to its coupling to vacuum fluctuations is in general
unobservable, as, practically, any phase variation is observed only
via some kind of interferometry between the involved state and
certain selected reference states which are both inseparably coupled
to vacuum. Nevertheless, if, somehow,
 vacuum fluctuations are modified, then the geometric phase of the
nonunitary evolution of an open system caused by its coupling to
vacuum may become potentially observable. The modification of vacuum
fluctuations induced by  the acceleration of a two-level atom, for
example, may provide such as a possibility,
 since, as is well-known, a uniformly accelerated observer perceives the Minkowski vacuum as a thermal
bath of Rindler particles~\cite{Unruh}. This is the so-called Unruh
effect. So, the phase variation due to the acceleration of an
two-level atom, which can in principle be  observed through
interference with an inertial atom,  may provide  evidence of the
Unruh effect which is deeply related to the Hawking radiation. In
this regard, let us note that many novel proposals have been
suggested to detect the Unruh effect and the Hawking radiation
 in analog systems~\cite{UH}. At this point, it may be worth
 pointing out that the Unruh effect is associated
 with quantization of the field in the Rindler accelerated frame.
 However, theoretical calculations performed from the  perspectives of  both the inertial frame and the Rindler accelerated frame
with the Unruh thermal bath usually produce the same result
 on physical observables~\cite{Matsas08}, as is the case in
the weak decay of a uniformly accelerated proton~\cite{Matsas01},
the bremsstrahlung effect associated with a uniformly accelerated
point charge~\cite{Higuchi92,Danilo}, and the spontaneous excitation
of a uniformly accelerated atom~\cite{ZhYu07}

Recently, Martin-Martinez et al.~\cite{Martin} have  considered the
possibility of using   geometric phase to detect the Unruh effect.
They examined an accelerated detector modeled by a harmonic
oscillator which couples only to a single-mode of a scalar field in
vacuum, and calculated the geometric phase acquired by the joint
state of the detector and  the field. As a result, cavities which
are leaky to a finite number of modes are essential for the
measurement of the acceleration influence in order to realize the
single mode coupling and avoid the problems in the Unruh effect
itself arising from introduction of boundaries. Such kind of
cavities seems to be a major challenge in experimental
implementation of their proposal. Here, we would like to consider a
more realistic case and propose using the geometric phase of
non-unitary evolution to detect the Unruh effect. We plan to study
an accelerated two-level system which couples to all vacuum modes of
electromagnetic (rather than scalar ) fields in a realistic
multipolar coupling scheme~\cite{CPP95}.  We treat the accelerated
two-level atom as an open system~\footnote{Let us note that the
theory of open quantum system  has  been fruitfully applied to
understand, from a different perspective,  the Unruh, Hawking and
Gibbons-Hawking effects, in Ref.~\cite{Benatti1}, \cite{yu3}
and~\cite{yu4}, respectively.} in a reservoir of fluctuating vacuum
electromagnetic fields and calculate the geometric phase of the
accelerated open system undergoing non-unitary evolution because of
the environment induced decoherence and dissipation. Since in our
study, the atom couples to all vacuum modes, no cavity is needed in
any experimental scheme to detect the phase. At this point, it is
worth noting that the quantum geometric phase of an open system
undergoing nonunitary evolution due to its coupling to a quantum
critical bath has recently been demonstrated using a NMR quantum
simulator~\cite{critical}.


Let us write the total Hamiltonian of the system (atom plus
reservoir) as $H=H_s+H_\phi+H'\;.$ Here $H_s$ is the Hamiltonian of
the atom, and, for simplicity, is taken to be $H_s={1\over
2}\,\hbar\omega_0\sigma_3,$ in which $\sigma_3$ is the Pauli matrix.
$\omega_0$ is the energy level spacing of the atom. $H_\phi$ is the
Hamiltonian of the free electromagnetic field, of which the details
are not needed here. The Hamiltonian that describes the interaction
between the atom and the  electromagnetic field in the multipolar
coupling scheme is given by $ H'(\tau)=-e\textbf{r} \cdot
\textbf{E}(x(\tau))=-e\sum_{mn}\textbf{r}_{mn}\cdot
\textbf{E}(x(\tau))\sigma_{mn}\;, $ where {\it e} is the electron
electric charge, $e\,\bf r$ the atomic electric dipole moment, and
${\bf E}(x)$ the electric field strength.

At the beginning, the whole system is characterized by the total density matrix $\rho_{tot}=\rho(0) \otimes |0\rangle\langle0|$, in which $\rho(0)$ is the initial reduced density matrix of the atom, and $|0\rangle$ is the vacuum state of the field. In the frame of the atom, the evolution in the proper time $\tau$ of the total density matrix $\rho_{tot}$ satisfies
\begin{equation}
\frac{\partial\rho_{tot}(\tau)}{\partial\tau}=-{i\/\hbar}[H,\rho_{tot}(\tau)]\;.
\end{equation}
We assume that the interaction between the atom and the field is
weak. In the limit of weak coupling, the evolution of the reduced
density matrix $\rho(\tau)$ can be written in the
Kossakowski-Lindblad form~\cite{Lindblad, pr5}
\begin{equation}\label{master}
{\partial\rho(\tau)\over \partial \tau}= -{i\/\hbar}\big[H_{\rm eff},\,
\rho(\tau)\big]
 + {\cal L}[\rho(\tau)]\ ,
\end{equation}
where
\begin{equation}
{\cal L}[\rho]={1\over2} \sum_{i,j=1}^3
a_{ij}\big[2\,\sigma_j\rho\,\sigma_i-\sigma_i\sigma_j\, \rho
-\rho\,\sigma_i\sigma_j\big]\ .
\end{equation}
The matrix $a_{ij}$ and the effective Hamiltonian $H_{\rm eff}$ are determined by the Fourier and Hilbert transforms of the field correlation functions
\begin{equation}
G^{+}(x-y)={e^2\/\hbar^2} \sum_{i,j=1}^3\langle +|r_i|-\rangle\langle -|r_j|+\rangle\,\langle0|E_i(x)E_j(y)|0 \rangle\;,
\end{equation}
which are defined as follows
\begin{equation}
{\cal G}(\lambda)=\int_{-\infty}^{\infty} d\tau \,
e^{i{\lambda}\tau}\, G^{+}\big(x(\tau)\big)\; ,
\quad\quad
{\cal K}(\lambda)=\frac{P}{\pi
i}\int_{-\infty}^{\infty} d\omega\ \frac{ {\cal G}(\omega)
}{\omega-\lambda} \;.
\end{equation}
Then the coefficients of the Kossakowski matrix $a_{ij}$ can be written as
\begin{equation}
a_{ij}=A\delta_{ij}-iB
\epsilon_{ijk}\delta_{k3}+C\delta_{i3}\delta_{j3}\;,
\end{equation}
in which
\begin{equation}\label{abc}
A=\frac{1}{4}[{\cal {G}}(\omega_0)+{\cal{G}}(-\omega_0)]\;,\;~~
B=\frac{1}{4}[{\cal {G}}(\omega_0)-{\cal{G}}(-\omega_0)]\;,~~
C=-A\;.
\end{equation}
The effective Hamiltonian $H_{\rm eff}$ contains a correction term,
the so-called Lamb shift, and one can show that it can be obtained
by replacing $\omega_0$ in $H_s$ with a renormalized energy level
spacing $\Omega$ as follows
\begin{equation}\label{heff}
H_{\rm eff}=\frac{1}{2}\hbar\Omega\sigma_3={\hbar\over 2}\{\omega_0+{i\/2}[{\cal
K}(-\omega_0)-{\cal K}(\omega_0)]\}\,\sigma_3\;.
\end{equation}

For convenience, let us express the density matrix $\rho$ in terms of the Pauli matrices,
\begin{equation}\label{density}
\rho({\tau})=\frac{1}{2}\bigg(1+\sum_{i=1}^{3}\rho_i({\tau})\sigma_i\bigg)\;.
\end{equation}
Plugging Eq.~(\ref{density}) into Eq.~(\ref{master})
%
and assuming that the initial state of the atom is
$|\psi(0)\rangle=\cos{\theta\/2}|+\rangle+\sin{\theta\/2}|-\rangle$,
 we can easily work out the time-dependent reduced density matrix
\begin{equation}\label{dens}
\rho(\tau)=\left(
\begin{array}{ccc}
e^{-4A\tau}\cos^2{\theta\/2}+{B-A\/2A}(e^{-4A\tau}-1) & {1\/2}e^{-2(2A+C)\tau-i\Omega\tau}\sin\theta\\ {1\/2}e^{-2(2A+C)\tau+i\Omega\tau}\sin\theta & 1-e^{-4A\tau}\cos^2{\theta\/2}-{B-A\/2A}(e^{-4A\tau}-1)
\end{array}\right)\;.
\end{equation}

The geometric phase for a mixed state undergoing nonunitary
evolution is given by~\cite{Tong}
\beq\label{gp} \gamma=\arg
\left( \sum\limits_{k=1}^N \sqrt{\lambda_k(0)\lambda_k(T)}\langle
\phi_k(0)|\phi_k(T)\rangle e^{-\int_0^T \langle \phi_k(\tau)|\dot
\phi_k(\tau) \rangle d\tau} \right)\;,
\eeq
where $\lambda_k(\tau)$
and $|\phi_k(\tau)\rangle$ are the eigenvalues and eigenvectors of
the reduced density matrix $\rho(\tau)$. In order to get the
geometric phase, we first calculate the eigenvalues of the density
matrix (\ref{dens}) to get
 \beq
\lambda_\pm(\tau)={1\/2}(1\pm\eta)\;,
 \eeq
 where $\eta=\sqrt{\rho_3^2+e^{-4(2A+C)\tau}\sin^2\theta}$
and $\rho_3=e^{-4A\tau}\cos\theta+{B\over A}(e^{-4A\tau}-1)$. It is
obvious that $\lambda_-(0)=0$.  As a result, the contribution comes
only from the eigenvector corresponding to $\lambda_+$
\beq
|\phi_+(\tau)\rangle=\sin{\theta_{\tau}\/2}|+\rangle+\cos{\theta_{\tau}\/2}e^{i\Omega\tau}|-\rangle\;,
\eeq
where \beq\label{tan}
\tan{\theta_{\tau}\/2}=\sqrt{\eta+\rho_3\/\eta-\rho_3}\;. \eeq The
geometric phase can be calculated directly using Eq.~(\ref{gp}) \beq\label{gp1}
\gamma=-\Omega\int_0^T\cos^2{\theta_{\tau}\/2}\,d\tau\;. \eeq

Let us now calculate the geometric phase of an two-level atom which
is uniformly accelerated, for example, in the $x$-direction.  The
trajectory of the atom is then described by
\begin{eqnarray}\label{traj}
t(\tau)={c\/a}\sinh{a\tau\/c},\ \ \
x(\tau)={c^2\/a}\cosh{a\tau\/c},\ \ \ y(\tau)=z(\tau)=0\;.
\end{eqnarray}
In order to get the explicit form of the geometric phase, we need  the field correlation functions, which can be worked out using the two point function of the electric field
\begin{eqnarray}\label{vac-green}
\langle E_i(x(\tau))E_j(x(\tau'))\rangle={\hbar c\/4\pi^2\varepsilon_0}(\partial_0\partial_0^\prime\delta_{ij}-\partial_i\partial_j^\prime)
{1\/|{\bf x-x^\prime}|^2-(c\,t-c\,t'-i\varepsilon)^2}\;.
\end{eqnarray}
The field correlation function for the trajectory (\ref{traj}) can
then be evaluated from (\ref{vac-green}) in the frame of the atom to
get
\begin{eqnarray}\label{wightman1}
G^+(x,x')={e^2|\langle -|{\bf r}|+\rangle|^2\/16\pi^2\varepsilon_0\hbar c^7} {a^4\/\sinh^4[{a\/2c}(\tau-\tau'-i\varepsilon)]}\;.
\end{eqnarray}
So, the Fourier transform of the field correlation function is
\begin{eqnarray}\label{fourier}
{\cal G}(\lambda)={\lambda^3\,e^2|\langle -|{\bf r}|+\rangle|^2\/6\pi\varepsilon_0\hbar c^3}\bigg(1+{a^2\/c^2\lambda^2}\bigg) \bigg(1+\coth{\pi c\lambda\/a}\bigg)\;.
\end{eqnarray}
Consequently, the coefficients of the Kossakowski matrix $a_{ij}$
and the effective level spacing of the atom are given by
\bea
  &&A_a=-C_a={1\/4}\gamma_0\,\bigg(1+{a^2\/c^2\omega_0^2}\bigg)\,\frac{e^{2\pi c\,\omega_0/a}+1}{e^{2\pi c\,\omega_0/a}-1}\;,
  \quad\quad\quad B_a={1\/4}\gamma_0\,\bigg(1+{a^2\/c^2\omega_0^2}\bigg)\;,\\
  &&\Omega_{a}=\omega_0+{\gamma_0P\/2\pi\omega_0^3}\int_0^\infty
d\omega\,\omega^3\bigg({1\/\omega+\omega_0}-{1\/\omega-\omega_0}\bigg)\bigg(1+{a^2\/c^2\omega^2}\bigg)\bigg(1+{2\/e^{2\pi
c\,\omega/a}-1}\bigg)\;,
\eea
where $\gamma_0=e^2|\langle -|{\bf r}|+\rangle|^2\,\omega_0^3/3\pi\varepsilon_0\hbar c^3$ is the
spontaneous emission rate.
Then the geometric phase can be obtained according to
\beq\label{acc}
\gamma_{a}=-\int_0^{T}{1\over2}\bigg(1-\frac{R-R\,e^{4A_a\tau}+\cos\theta}{\sqrt{e^{4A_a\tau}\sin^2\theta
+(R-R\,e^{4A_a\tau}+\cos\theta)^2}}\bigg)\,\Omega_a\,d\tau\;, \eeq
where $R=B_a/A_a$. So, the phase accumulates as the system evolves,
although the accumulation with time is not linear as in the unitary
evolution case. For a single period of evolution, the result of this
integral can be expressed as \beq
\gamma_{a}={\Omega_a\/\omega_0}\big[F(2\pi)-F(0)\big]\;, \eeq where
the function $F(\varphi)$ is defined as \bea
F(\varphi)&=&-{1\/2}\,\varphi-{1\/8A_a}\ln \left({\frac {1-Q^2-R^2+2R^2\,e^{4A_a\varphi/\omega_0}}{2R}}+S(\varphi) \right)\nn\\
&&-{1\/8A_a}\text{sgn}(Q)\ln \left({ {1-Q^2-{R}^{2}
+2Q^2e^{-4A_a\varphi/\omega_0}}+2\,|Q|S(\varphi)\,{e^{-4A_a\varphi/\omega_0)}}}
\right)\;, \eea in which $S(\varphi)=\sqrt
{R^2\,e^{8A_a\varphi/\omega_0}+(1-Q^2-{R}^{2})
{e^{4A_a\varphi/\omega_0}}+Q^2}\;$, $Q=R+\cos\theta$ and
$\text{sgn}(Q)$ is the standard sign function. For small
$\gamma_0/\omega_0$, which is generally true as we will see later,
we can perform a series expansion to the result. For a single
quasi-cycle, we find, to the first order~\footnote{Here we have
omitted the Lamb shift terms, since it is obvious that these terms
contain a factor $\gamma_0/\omega_0$ and they will only contribute
to the phase at the second and higher orders of
$\gamma_0/\omega_0$.}, \beq
\gamma_{a}\approx-\pi(1-\cos\theta)-\pi^2{\gamma_0\/2\omega_0}\sin^2\theta
\bigg(1+{a^2\/c^2\omega_0^2}\bigg)\bigg(2+\cos\theta+\frac{2}{e^{2\pi
c\,\omega_0 /a}-1}\cos\theta\bigg)\;. \eeq The first term
$-\pi(1-\cos\theta)$ in the above equation is the geometric phase we
would have obtained if the system were isolated from the
environment, and the second term is the correction induced by the
interaction between the accelerated atom and the environment. The
geometric phase contains a term proportional to $a^2$ apart from the
usual thermal term with a Planckian factor, and this term becomes
appreciable  when the acceleration is of the order of $c\,\omega_0$,
thus it enhances the accumulation of the geometric phase in contrast
with the scalar field case where this term is absent. Let us note
here that similar $a^2$ terms also appear in  the studies of the
energy shift~\cite{Passante} and the spontaneous
excitation~\cite{zhu} of an accelerated atom  once the scalar filed
is replaced by  the  electromagnetic field. In the limit of
$a\rightarrow0$, which corresponds to the case of an inertial atom,
there is still a correction, which comes from the zero point
fluctuations of  the Minkowski vacuum. The explicit form of this
term reads \beq\label{vac}
\gamma_{I}\approx-\pi(1-\cos\theta)-\pi^2{\gamma_0\/2\omega_0}(2+\cos\theta)\sin^2\theta\;.
\eeq This correction is exactly the same as the one in
Ref.~\cite{chen}, which is obtained by assuming an environment with
a Lorentzian spectral density, and is very similar to the result in
Ref.~\cite{Marzlin} derived from a different model. Thus the
correction to the geometric phase purely due to the acceleration is
\beq\label{GPdiff} \delta_a=\gamma_a-\gamma_I \approx
-\pi^2{\gamma_0\/2\omega_0}\bigg[{a^2\/c^2\omega_0^2}(2+\cos\theta)+
\bigg(1+{a^2\/c^2\omega_0^2}\bigg)\frac{2}{e^{2\pi
c\,\omega_0/a}-1}\cos\theta \bigg]\sin^2\theta\;.
 \eeq
This reveals that the geometric phase difference between the
accelerated and inertial atoms depends on the properties of the atom
(transition frequency $\omega_0$ and the spontaneous emission rate
$\gamma_0$), the initial state (angle $\theta$), and the
acceleration $a$. If we assume that $|\langle -|{\bf r}|+\rangle|$
is of the order of the Bohr radius $a_0$, and $\omega_0$  of the
order of $E_0/\hbar$, where $E_0=-e^2/8\pi\varepsilon_0a_0$ is the
energy of the ground-state, then $\gamma_0/\omega_0$ is of the order
of $10^{-6}$. For a given initial state, the phase difference
increases with the acceleration, and it becomes significant when the
acceleration is of the order $c\,\omega_0$. The initial state of the
atom, i.e., the initial angle $\theta$ in the Bloch sphere
representation, also plays an important role. When $\theta=0$ and
$\theta=\pi$, which corresponds to an initial excited state  and an
ground state respectively, the phase difference vanishes, whereas it
reaches its maximum in the regime near $\theta=\pi/2$. For a typical
transition frequency of the hydrogen atom, i.e.,
$\omega_0\sim10^{15}~{\rm s}^{-1}$, the acceleration needed to
observe this effect is of the order of $10^{23}~{\rm m/s^2}$, which
is extremely high. However, if we consider two-level systems with
lower frequency, the acceleration needed can be smaller. If we
choose transition frequencies of the atom  in the microwave regime,
for example, $\omega_0=2.0\times10^{9}~{\rm s}^{-1}$, which is
physically accessible~\cite{freq1,freq2}, then, for
$a=4\,c\,\omega_0=2.4\times10^{18}~{\rm m/s^2}$, the phase
difference can reach $1.6\times10^{-4}~{\rm rad}$ after a single
period of evolution, which may be within the current experimental
precision.


The geometric phase discussed above, therefore, may be detected with
an atom interferometer. One first prepares the two-level atom in a
superposition of upper and lower states in a Ramsey zone. In one arm
of the interferometer the atoms move inertially, and in the other
arm the atoms are accelerated. An interferometric measurement is
taken when the atoms in the two arms meet. Here, let us recall that
our calculations  of the geometric phase are based on the comoving
frame of reference of the atom. So, in the example we consider,
according to Eq.~(\ref{traj}), a single period for the accelerated
atom in the comoving frame $T=2\pi/\omega_0\sim
3.1\times10^{-9}~{\rm s}$ would transfer to a time interval of
$5.1~{\rm s}$ in the laboratory frame. Thus, one should prepare an
inertial atom which moves fast enough so that a single period of
time in its own frame also transfers to the same amount of time in
the laboratory frame when  the interference experiment is performed.
A tricky point is whether the field to accelerate the atoms will
change the structure of a real atom or even ionize it.

 Another delicate issue in experimental implementation is how to cancel the dynamical
phase that the atoms acquire. For systems under nonunitary evolution
like what we are considering here, the removal of the dynamical
phase from the total phase is a subtle issue~\cite{remove}. However,
since our purpose is to detect the Unruh effect associated with the
acceleration of the atom, we do not really need a complete
cancellation of the dynamical phase. Instead, we may choose slightly
different paths to control the relative dynamical phase to be much
smaller than the geometric phase acquired in one period, so the
result is effectively dominated by the geometric phase difference
between the accelerated and inertial atoms.





We would like to thank the Kavli Institute for Theoretical Physics China for hospitality where this work
was finalized. This work was supported in part by the NSFC under Grants No. 11075083 and No. 10935013, the Zhejiang Provincial Natural Science Foundation of China under Grant No. Z6100077, the National Basic Research Program of China under Grant No. 2010CB832803, the Program for Changjiang Scholars and Innovative Research Team in University (PCSIRT,  No. IRT0964), and the Hunan Provincial Natural Science Foundation of
China under Grant No. 11JJ7001.




\begin{thebibliography}{00}

\bibitem{Pancharatnam}
S. Pancharatnam, Proc. Indian Acad. Sci. A {\bf 44}, 247 (1956).

\bibitem{berry}
 M. V. Berry, Proc. R. Soc. Lond. A {\bf 392}, 45 (1984).

\bibitem{nonadiabatic} Y. Aharonov and J. Anandan, Phys. Rev. Lett.
{\bf 58}, 1593 (1987).

\bibitem{noncyclic} J. Samuel and R. Bhandari, Phys. Rev. Lett. {\bf
60}, 2339 (1988).

\bibitem{GPbook}
{\it Geometric Phases in Physics}, edited by A. Shapere and
F.Wilczek (World Scientific, Singapore, 1989).

\bibitem{molecular}
S. R. Jain and A. K. Pati, Phys. Rev. Lett. {\bf 80}, 650 (1998).

\bibitem{solid}
D. Xiao, M. Chang and Q. Niu, Rev. Mod. Phys. {\bf 82}, 1959 (2010).

\bibitem{fault-tolerant}
J. A. Jones, V. Vedral, A. Ekert, and G. Castagnoli, Nature {\bf
403}, 869 (2000).

\bibitem{Uhlmann}
A. Uhlmann, Rep. Math. Phys. {\bf 24}, 229 (1986).

\bibitem{Sjoqvist}
E. Sj\"{o}qvist, A. K. Pati, A. Ekert, J. S. Anandan, M. Ericsson,
D. K. L. Oi, and V. Vedral, Phys. Rev. Lett. {\bf 85}, 2845 (2000).

\bibitem{Singh}
K. Singh, D. M. Tong, K. Basu, J. Chen, and J. Du, Phys. Rev. A {\bf
67}, 032106 (2003).

\bibitem{Tong}
D. M. Tong, E. Sj\"{o}qvist, L. C. Kwek, and C. H. Oh, Phys. Rev.
Lett. \textbf{93}, 080405 (2004).

\bibitem{Wang}
Z. S. Wang, L. C. Lwek, C. H. Lai, and C. H. Oh, Europhys. Lett.
{\bf 74}, 958 (2006).

\bibitem{Du}
J. Du, P. Zou, M. Shi, L. C. Kwek, J. Pan, C. H. Oh, A. Ekert, D. K.
L. Oi, and M. Ericsson, Phys. Rev. Lett. {\bf 91}, 100403 (2003).

\bibitem{Marie}
M. Ericsson, D. Achilles, J. T. Barreiro, D. Branning, N. A. Peters,
and P. G. Kwiat, Phys. Rev. Lett. {\bf 94}, 050401 (2005).

\bibitem{Carollo}
A. Carollo, I. Fuentes-Guridi, M. F. Santos, and V. Vedral, Phys.
Rev. Lett. {\bf 90}, 160402 (2003); ibid. {\bf 92}, 020402 (2004).

\bibitem{Rezakhani}
A. T. Rezakhani and P. Zanardi, Phys. Rev. A {\bf 73}, 052117
(2006).

\bibitem{Lombardo}
F. C. Lombardo and P. I. Villar, Phys. Rev. A {\bf 74}, 042311
(2006).

\bibitem{chen}
J. Chen, J. An, Q. Tong, H. Luo, and C. H. Oh, Phys. Rev. A {\bf
81}, 022120 (2010).


\bibitem{Unruh}
W. G. Unruh, Phys. Rev. D {\bf 14}, 870 (1976).

\bibitem{UH}
W. G. Unruh, Phys. Rev. Lett. {\bf 46}, 1351 (1981);  L. J. Garay,
J. R. Anglin, J. I. Cirac, and P. Zoller, Phys. Rev. Lett. {\bf 85},
4643 (2000); U. Leonhardt, Nature {\bf 415}, 406 (2002); T. G.
Philbin and et al., Science {\bf 319}, 1367 (2008);  P. D. Nation,
M. P. Blencowe, A. J. Rimberg, and E. Buks, Phys. Rev. Lett. {\bf
103}, 087004 (2009).

\bibitem{Matsas08} For a comprehensive  review, see for example, L. C. B. Crispino, A. Higuchi, and G. E. A. Matsas,
Rev. Mod. Phys. {\bf80}, 787 (2008).

\bibitem{Matsas01} G. E. A. Matsas and D. A. T. Vanzella, Phys. Rev.
Lett {\bf87}, 151301 (2001).
\bibitem{Higuchi92} A. Higuchi, G. E. A. Matsas and D. Sudarsky,
Phys. Rev. D {\bf45}, R3308 (1992); Phys. Rev. D {\bf46}, 3450
(1992).
\bibitem{Danilo} Danilo T. Alves and Lu\'{\i}s C. B. Crispino, Phys. Rev.
D {\bf70}, 107703 (2004).

\bibitem{ZhYu07}Z. Zhu and H. Yu, Phys. Lett. B {\bf 645},
459(2007); Chin. Phys. Lett. {\bf 25}, 1575(2008).

\bibitem{Martin}
E. Martin-Martinez, I. Fuentes and R. B. Mann, Phys. Rev. Lett. {\bf 107}, 131301 (2011).

\bibitem{CPP95} G. Compagno, R. Passante, and F. Persico, {\it
Atom-Field Interactions and Dressed Atoms} (Cambridge University
Press, Cambridge, England, 1995).

\bibitem{Benatti1}
F. Benatti and  R. Floreanini , Phys. Rev. A {\bf 70}, 012112
(2004).

\bibitem{yu3}
H. Yu and J. Zhang, Phys. Rev. D {\bf 77}, 024031 (2007).


\bibitem{yu4}
H. Yu, Phys. Rev. Lett. {\bf 106}, 061101 (2011).


\bibitem{critical}
F. M. Cucchietti, J.-F. Zhang, F. C. Lombardo, P. I. Villar, and R.
Laflamme, Phys. Rev. Lett. {\bf 105}, 240406 (2010).

\bibitem{Lindblad}
V. Gorini, A. Kossakowski, and E. C. G. Surdarshan, J. Math. Phys.
{\bf 17}, 821 (1976); G. Lindblad, Commun. Math. Phys. {\bf 48}, 119
(1976).



\bibitem{pr5}
F. Benatti, R. Floreanini and M. Piani, Phys. Rev. Lett. {\bf 91},
 070402  (2003).

\bibitem{Passante}
R. Passante, Phys. Rev. A {\bf 57}, 1950 (1998).

\bibitem{zhu}
Z. Zhu, H. Yu and S. Lu, Phys. Rev. D {\bf 73}, 107501 (2006).

\bibitem{Marzlin}
K.-P. Marzlin, S. Ghose, and B. C. Sanders, Phys. Rev. Lett. {\bf 93}, 260402  (2004).

\bibitem{freq1}
J. M. Raimond, M. Brune, and S. Haroche, Rev. Mod.
Phys. {\bf 73}, 565 (2001).

\bibitem{freq2}
M. O. Scully, V.V. Kocharovsky, A. Belyanin, E. Fry, and
F. Capasso, Phys. Rev. Lett. {\bf 91}, 243004 (2003).


\bibitem{remove}
E. Sj\"{o}qvist, Physics 1, 35 (2008).







\end{thebibliography}
\end{document}